# Phase-Induced Frequency Conversion and Doppler Effect with Time-Modulated Metasurfaces

Davide Ramaccia, *Senior Member, IEEE*, Dimitrios L. Sounas, *Senior Member, IEEE,*
Andrea Alù, *Fellow, IEEE,* Alessandro Toscano, *Senior Member, IEEE,* and Filiberto Bilotti, *Fellow, IEEE*

*Abstract*— Metasurfaces consisting of electrically thin and densely packed planar arrays of subwavelength elements enable an unprecedented control of the impinging electromagnetic fields. Spatially modulated metasurfaces can efficiently tailor the spatial distribution of these fields with great flexibility. Similarly, time modulated metasurfaces can be successfully used to manipulate the frequency content and time variations of the impinging field. In this paper, we present time-modulated reflective metasurfaces that cause a frequency shift to the impinging radiation, thus realizing an artificial Doppler effect in a non-moving electrically thin structure. Starting from the theoretical analysis, we analytically derive the required time modulation of the surface admittance to achieve this effect, and present a realistic time-varying structure, based on a properly designed and dynamically tuned high-impedance surface. It is analytically and numerically demonstrated that the field emerging from the metasurface is up-/down-converted in frequency according to the modulation profile of the metasurface. The proposed metasurface concept, enabling a frequency modulation of the electromagnetic field "on-the-fly", may find application in telecommunication, radar, and sensing scenarios.

## I. Introduction

METASURFACES have produced a dramatic impact on electromagnetic engineering thanks to their ability to manipulate the wavefront of the reflected and transmitted waves at will [1], [2]. They are typically made of a planar dense array of resonant or nearly resonant subwavelength elements, whose properties are determined by both the specific features of the subwavelength elements and their coupling, determined by the array geometry and possible presence of a supporting substrate [3]. By properly designing these subwavelength elements, it is possible to engineer the phase discontinuity across the metasurface according to the generalized Snell's laws [4], and an additional tangential momentum can be imparted to the incident field in order to redirect the scattered field in one or more desired directions. This feature has been used for numerous applications, such as beam steering [5]–[11], focusing [12]–[14], filtering [15]–[17], holography [18]–[20], just to name a few. Assuming that the metasurface properties do not change in time and its response is linear, the structure is able to manipulate only the wave-vector spectrum of the incident field, leaving the frequency spectrum unaltered.

The control of the metamaterial/metasurface response in time has been first attempted through tunability. However, tunable metamaterials and metasurfaces have been mostly studied in the quasi-static case, where temporal variations can be treated as a tuning parameter of the structure's static response. Only recently the effects of time modulation beyond the quasi-static regime has been investigated [21], [22]. It has been demonstrated that modulation in time enables several exotic space–time scattering phenomena never observed before. Among them, the possibility to break the reciprocity constraint and generate/control frequency harmonics in the scattered field represents the most appealing functionality [23]–[30]. Space-time-varying metamaterials exhibit dielectric properties modulated in both space and time [31]–[34]. The refractive index changes periodically across space and time, and it can be made to appear as moving in one specific direction with a certain velocity, thereby allowing breaking reciprocity [21], [22], [33], [35], [36].

As it happens in any actual moving medium, the electromagnetic wave interacts with such a material differently depending on the illumination direction. Thanks to time-reversal symmetry breaking induced by the apparent motion of the medium, this effect has been exploited to conceive several non-reciprocal components and devices, which behave similarly to their conventional counterparts based on magnets.

Manuscript received April 29, 2019.

Davide Ramaccia, Filiberto Bilotti, and Alessandro Toscano are with the Department of Engineering at ROMA TRE University, Via Vito Volterra 62, 00146 Rome, Italy (e-mail: davide.ramaccia-, alessandro.toscano-, filiberto.bilotti@uniroma3.it).

Dimitrios L. Sounas is with the Department of Electrical and Computer Engineering, Wayne State University, Detroit, MI 48202, USA (e-mail: dsounas@wayne.edu).

Andrea Alù is with the Photonics Initiative, Advanced Science Research Center, the Physics Program, Graduate Center, and the Department of Electrical Engineering, City College, all at the City University of New York, New York, NY 10031, USA (e-mail: aalu@gc.cuny.edu).

The work has been developed in the framework of the activities of the research contract MANTLES, funded by the Italian Ministry of Education, University and Research as a PRIN 2017 project (protocol number 2017BHFZKH).

Some relevant examples are magnet-less isolators [21], [22], circulators [37]–[39], and non-reciprocal antennas [40]–[44]. Moreover, the ability of space-time modulated metamaterials to shift the frequency spectrum of the impinging field according to the modulation scheme has been exploited to conceive a Doppler cloak, i.e., a space-time modulated cover that compensates for the Doppler shift of a moving object, thus restoring the source frequency in reflection as if the covered scatterer were at rest [45]–[47]. So far, most practical realizations of time-modulated structures have been for guided waves. On the contrary, only few realistic metasurfaces with time-varying surface impedance have been proposed [23], [26], [30], [48], [49], even though several theoretical works on the topic have envisioned interesting applications (Ref. [24], [25], [50], [51] and references within).

In this paper, we introduce a simple and effective design approach for time-modulated reflective metasurfaces, able to introduce either a red or blue frequency shift to an incident electromagnetic wave, realizing, thus, an artificial Doppler shift through a non-moving electrically thin reflective surface. Starting from a theoretical analysis, we derive the required time-domain surface admittance of the metasurface and present the design of a realistic time-varying metasurface, consisting of a bi-dimensional array of metal-backed mushroom-like elements, which is dynamically tuned through a set of varactors. It is analytically and numerically demonstrated that the reflected field is frequency shifted by just electrically controlling the phase delay of the reflection coefficient in real time.

This manuscript is organized as follows. In Section II, we present the theoretical analysis of the scattering from a reflective metasurface, whose surface admittance $Y_s$ is time-modulated. We demonstrate that the desired frequency shift is achieved if the metasurface can be tuned from a perfect electric to a perfect magnetic conductor response along one modulation period. In Section III, we study the response of a properly designed polarization-insensitive tunable metasurface, which, at the design frequency, appears as either a perfect electric or a perfect magnetic conductor, depending on the applied electric voltage at the varactor terminals. In Section IV, the frequency conversion of the time-varying metasurface is verified. First, we design the time-varying driving voltage signal to satisfy the analytical conditions for achieving an up- and down-conversion of the illuminating plane wave, and, then, a full-wave time-domain simulation is performed, demonstrating that the frequency of the reflected electromagnetic wave can be actually red- or blue-shifted with respect to the incident one by controlling the electric driving voltage signal of the varactors. Finally, in Section V, some conclusions are drawn.

## II. MODULATION-INDUCED ARTIFICIAL DOPPLER FREQUENCY SHIFT

The aim of this Section is to define the design rules for a general time-modulated fully reflective metasurface,

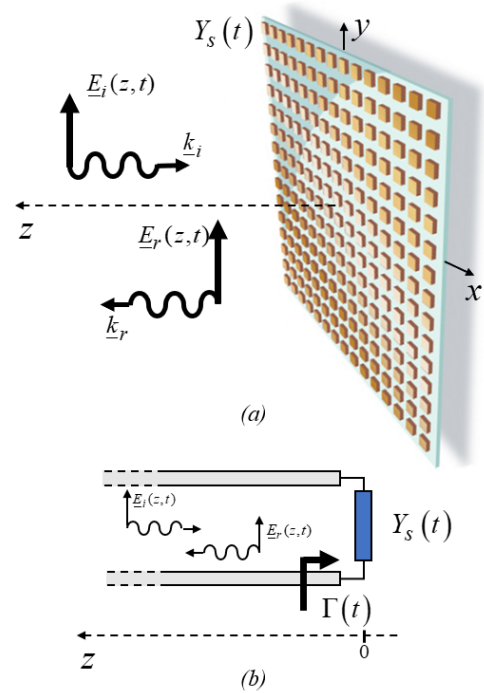

Fig. 1. (a) A metasurface with a time-varying surface admittance $Y_s(t)$ is illuminated by a normally impinging electromagnetic field at frequency $\omega_i = c_0 k_i$ and propagating in negative z-direction. A field at different frequency $\omega_r = c_0 k_r$ is reflected back. (b) Equivalent transmission line model of the system reported in Fig. 1a.

exploiting the analytical expression of the reflection coefficient. In order to gain insight into the scattering of the electromagnetic field from a periodically time-varying reflective metasurface, we limit our analysis to the case of normal incidence. This allows deriving the general time-dependent relations between the incident and the reflected fields, and between the reflection coefficient and the surface admittance, which are the fundamental quantities to design the needed temporal profile that imparts a controlled frequency shift to the incident electromagnetic wave.

Let us consider a planar metasurface consisting of a dense periodic array of subwavelength elements, whose electromagnetic response varies periodically in time. If the size of the elements is much smaller than the operating wavelength, the array can be modeled through an effective surface admittance $Y_s$. In general, metasurfaces exhibit a frequency dispersive surface admittance, i.e. $Y_s = Y_s(\omega)$. However, if the metasurface exhibits a slow frequency dispersion around the illumination frequency, and the modulation-induced frequency shift is much smaller than the illumination frequency, we can assume that the metasurface exhibits the same surface properties for both the incident and reflected fields. This allows to simplify the mathematical analysis, neglecting the modulation-induced dispersion effects, and considering the surface admittance as only a function of time, i.e. $Y_s = Y_s(t)$.

If the metasurface is fully reflective, no transmitted field is present beyond the metasurface. Therefore, the total

electromagnetic field in time domain is a superposition of the normally incident and reflected fields:

$$\underline{E}_i(z,t) = \text{Re}\left\{\underline{E}_{0,i} e^{-jk_i z} e^{j\omega_i t}\right\}$$
$$\underline{E}_r(z,t) = \text{Re}\left\{\underline{E}_{0,r} e^{+jk_r z} e^{j\omega_r t}\right\} \quad (1)$$

where $\underline{E}_{0,i(r)}$, $k_{i(r)}$, and $\omega_{i(r)}$ are the complex electric field amplitude, free-space wavevector and angular frequency, respectively, and $\text{Re}\{\bullet\}$ denotes the real part of the complex quantity in the brackets. The entire system can be simply modeled though its equivalent transmission line model shown in Fig. 1b. The incident and reflected fields are related to each other by the time-varying complex reflection coefficient:

$$\Gamma(t) = \frac{Y_0 - Y_s(t)}{Y_0 + Y_s(t)} \quad (2)$$

where $Y_0 = 1/120\pi$ is the free-space admittance.

At the metasurface location, *i.e.* $z=0$, the incident and reflected electric fields in time domain are related as

$$\underline{E}_r(z=0,t) = \Gamma(t) \cdot \underline{E}_i(z=0,t) \quad (3)$$

and the corresponding equation in the angular frequency domain can be derived by taking the Fourier transform of both sides:

$$\underline{E}_r(\omega) = \underline{\Gamma}(\omega) * \underline{E}_i(\omega) = \int \underline{\Gamma}(\omega-\omega')\underline{E}_i(\omega')d\omega' \quad (4)$$

where $\underline{E}_i(\omega), \underline{E}_r(\omega)$ are the complex amplitudes of the incident and reflected propagating plane waves in the frequency domain, respectively, and the symbol "$*$" denotes convolution. Equation (4) implies that, whatever the illuminating frequency $\omega_i$ is, the input spectrum will be converted into a spectrum of output frequencies, according to the frequency spectrum of the reflection coefficient.

The time varying profile of the reflection coefficient $\Gamma(t)$ forced by the periodic modulation of the surface admittance $Y_s(t)$ can be expanded in a Fourier series as

$$\underline{\underline{\Gamma}}(t) = \sum_n \underline{\underline{\Gamma}}^n(\omega_i) e^{jn\omega_m t} \quad (5)$$

where $\omega_m$ is the angular modulation frequency. The Fourier transform of (5) takes the form

$$\underline{\underline{\Gamma}}(\omega) = \sum_n \underline{\underline{\Gamma}}^n(\omega)\delta(\omega - n\omega_m). \quad (6)$$

Therefore, inserting (6) into (4) and assuming that the incident field is monochromatic, we obtain the generated frequency spectrum of the reflected electromagnetic field:

$$\underline{E}_r(\omega) = \sum_n \underline{\underline{\Gamma}}^n(\omega_i)\underline{E}_{0,i}\delta(\omega-\omega_i + n\omega_m), \quad (7)$$

Equation (7) is a discrete spectrum consisting of the illumination frequency $\omega_i$ up- and down-modulated by the time modulation frequency $\omega_m$.

Equations (1)-(7) can now be used to derive the necessary temporal profile of the surface impedance to achieve a perfect red- or blue-shift of the incident monochromatic plane wave at frequency $\omega_i$ to the reflected one at frequency $\omega_r$. From (3) and (1) the reflection coefficient can be written as

$$\Gamma(\omega_i, t) = \frac{\underline{E}_r(z=0,t)}{\underline{E}_i(z=0,t)} = \frac{\underline{E}_{0,r}}{\underline{E}_{0,i}} e^{j(\omega_r - \omega_i)t}. \quad (8)$$

In the lossless case and for very small difference between incident and reflected frequencies, the amplitude of the reflection coefficient approaches unity according to the Manley-Rowe relations [52], [53]. Therefore, the reflection coefficient in Eq. (8) turns to be a simple time-varying phase, spanning the $2\pi$ range with angular frequency $\omega_r - \omega_i$. Comparing (8) with (5) for *n=1*, we note that such a frequency corresponds to the modulation frequency $\omega_m$ of the metasurface, *i.e.* $\omega_m = |\omega_r - \omega_i|$. This result is in line with the frequency shift resulting from the Doppler effect in presence of a moving reflector [34]. Indeed, let us consider a perfect reflector illuminated by a normal plane wave. The reflector is now moved away from the source with a uniform velocity *v* in the same propagation direction as the illuminating wave. As time goes on, the distance between the source and the reflector increases, while the phase of the reflection coefficient decreases, *i.e.* the phase delay increases, by a quantity $\Delta\varphi = -4\pi vt/\lambda_0$, where $\lambda_0$ is the free-space wavelength of the source, whereas the amplitude of the reflection coefficient is still unitary, as in (8). Here, the reflector is stationary, but the continuous reduction of reflection phase can emulate the motion of the reflector away from the source and, consequently, inducing an artificial Doppler shift. Similarly, a reflector approaching the source can be emulated by continuously increasing of the reflection phase.

The required surface admittance $Y_s(t)$ can be found by combining eqs. (2) and (8):

$$\frac{Y_s(t)}{Y_0} = -j\tan\left[\frac{1}{2}\omega_m t\right] \quad (9)$$

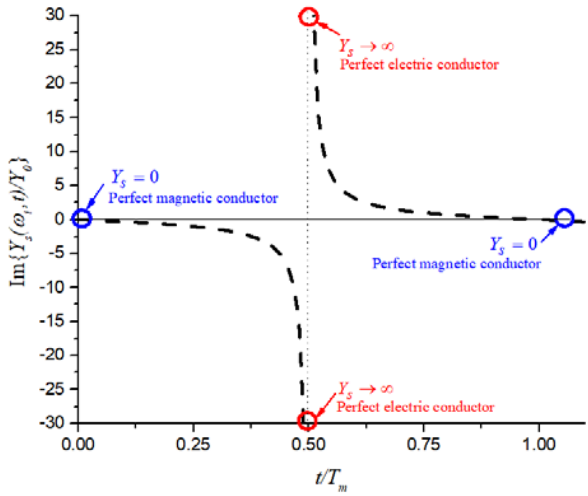

Fig. 2. Time-dependent profile of normalized surface admittance $Y_s(t)/Y_0$ as a function of normalized time with respect to one modulation period $T_m$.

As expected, if the system is lossless, the metasurface admittance is imaginary with an inductive or capacitive behavior according to the sign of the tangent function. The normalized surface admittance $Y_s/Y_0$ over a modulation period $T_m = 2\pi/\omega_m$ is shown in Fig. 2. For each modulation period, the surface admittance assumes all possible values following a tangent function profile, spanning from $Y_s = 0$ to $Y_s \to \pm\infty$. In particular, these two extreme values correspond to the case of a perfect magnetic and a perfect electric conductor, respectively. It is clear that such a metasurface belongs to the general class of meta-mirrors proposed in [14]. A meta-mirror consists of an array of electrically small resonant bi-anisotropic particles that can be designed to emulate the response of a perfect reflector with arbitrary reflection phase, including a perfect electric and magnetic conductors. However, the dynamic control of a bi-anisotropic particle is not an easy task, because the electric and magnetic responses should be simultaneously controlled in an independent way to modulate the phase of the reflection coefficient, while the zero-transmission condition holds. An attempt to achieve an electric control of the electric and magnetic polarizabilities independently has been recently proposed in [30], where, however, the complex biasing network strongly limits its implementation in a bi-dimensional structure.

An easier path to achieve a perfect reflecting surface whose response can be both a perfect electric and a perfect magnetic conductor is to use a *high-impedance surface* (HIS) [54]. HISs are electrically thin metallic structures forbidding the flow of alternating currents within a given frequency band, as for a perfect magnetic conductor. Outside such a frequency band, the surface behaves as a conventional perfect electric conductor. The dynamic tunability of the operating frequency band has been deeply investigated in the past [55] and it has been demonstrated that the electromagnetic response from a perfect electric to a perfect magnetic conductor can be controlled by a set of reverse-biased varactors. In the next Sections, we exploit the dynamic tunability of a reflective metasurface based on HIS to achieve the desired time-varying reflection coefficient and the equivalent input surface impedance described by eqs. (8) and (9), respectively.

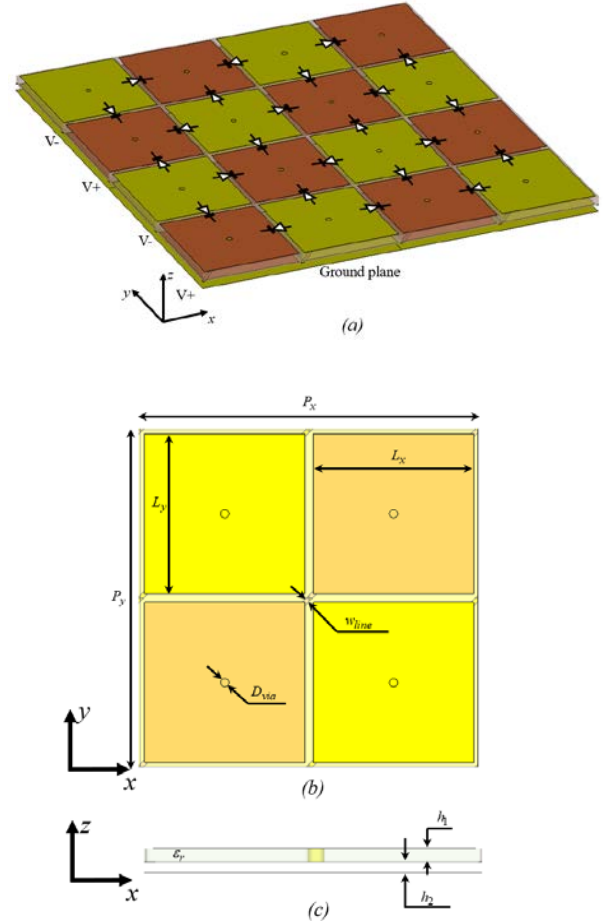

Fig. 3. (a) Prospective view, (b) top view and (c) side view with dimensions of the time-varying metasurface. Dimensions of the unit cell and patches are: $P_x = 80 mm$, $P_y = 80 mm$, $L_x = 38 mm$, $L_y = 38 mm$.

## III. GEOMETRY AND TUNABILITY OF THE TIME-VARYING METASURFACE

In this Section, we present the geometry of the time-modulated metasurface and its tuning scheme that allows us to dynamically change its response from a perfect electric to a perfect magnetic conductor, and vice-versa. The tunability is achieved using a set of electrically controlled varactors that change in real-time the frequency response of the metasurface, *i.e.* its complex reflection coefficient.

### A. Geometry

The proposed metasurface is shown in Fig. 3. It consists of five layers: the first three are used for realizing an array of mirrored mushrooms, the fourth is thin foam sheet, and the last layer is a metallic ground plane that enforces the zero transmission (see Fig. 3a). The mirrored mushrooms are composed by two metallic copper patches printed on the two sides of a dielectric substrate and connected by a via. The dielectric substrate is a standard lossy FR-4 with permittivity $\varepsilon_r = 4.3$, dissipation factor $\tan\delta = 0.025$ @ $f = 10$ GHz, and height $h_1 = 1.6 mm$. Four mirrored mushrooms realize a unit cell of the metasurface as shown in Fig. 3(b). The four patches are electrically connected to each other in pairs though a biasing line of width $w_{line} = 1$ mm, printed on the bottom layer of the FR-4 dielectric substrate and connecting the bottom patches. Thanks to the presence of the metallic via with diameter $D_{via} = 1$ mm, the bottom and top patches are electrically connected to each other. The biasing lines run at 45° with respect to the unit cell main axes in order to realize a chessboard electric potential level between all the patches in $x$- and $y$- directions. The different electric potential level of the patches is represented in Fig. 3 by using a different color for the patch, *i.e.* orange and yellow, according to the electric voltage applied through the biasing lines. Such a biasing network ensures that all the patches can apply simultaneously a reverse voltage bias to all the four varactors applied to their edges, as shown in Fig. 3(a). In this way, the metasurface can operate for both polarization of the normally impinging plane wave. The metallic ground plane is placed below the dielectric bottom layer, separated by a foam sheet of height $h_2 = 1.25$ mm. Finally, the unit cell and metallic patches of the *unloaded* metasurface have been designed to resonate at around 1.75 GHz. The dimensions of the unit cell and patches are reported in the caption of Fig. 3.

### B. Loading circuit and tunability

The metasurface is loaded with a set of varactors controlled by the voltage signal, applied at the terminals $V^+/V^-$ (Fig. 3). The varactor selected for the proposed metasurface is the *varactor BB535*, produced by Infineon [56]. It exhibits a diode capacitance $C_T$ in the range 2.1-18.7 pF for a corresponding applied reverse biasing voltage $V_{bias} = V^+ - V^-$ from 28 V to 1 V. The varactor is available in a package SOD323 (Fig. 4a), which adds both inductive and capacitive parasitic elements due to the metallic package and terminals.

In Fig. 4b, we report the complete loading circuit of our metasurface, where the varactor is modeled though its complete SPICE model and the parasitic reactive elements $L_1 = L_2 = 0.55$ nH, $L_3 = 0.67$ nH, $C_1 = 110$ fF of the package are taken into account. The polarization network consists of a dc and high-frequency filters through two series capacitors $C_{dc} = 1$ nF, blocking the dc voltage $V_{bias}$, and two choke inductors $L_{chk} = 1$ nH, acting as a low-pass filters rejecting the high frequency of the exciting electromagnetic field impinging on the metasurface. The terminals A-A' in Fig. 4b are connected to the edges of two adjacent patches on the metasurface.

The electromagnetic response of the metasurface has been numerically evaluated in CST Microwave Studio in co-simulation with CST Design Studio [57], where the circuit of Fig. 4b has been implemented. The amplitude and phase of the reflection coefficient at frequency $f_0 = 1.5$ GHz as a function of the biasing voltage $V_{bias}$ are reported in Fig. 4c for both horizontal (x-)polarization and vertical (y-)polarization. It is clear that, according to the applied voltage, the metasurface exhibits an electromagnetic response similar to a perfect electric and magnetic conductor when the phase approaches ±180° and 0°, respectively. The amplitude of the reflection coefficient is not unitary when a zero phase is approached, indicating some loss around this region. This is because the metasurface is in its resonant state for exhibiting a zero phase of the reflection coefficient and, consequently, part of the impinging energy is dissipated within the lossy dielectrics and metals of the structure, and the parasitic resistances of the varactors. However, as we will demonstrate in the next Section, such an attenuation does not affect the ability of the metasurface to frequency shift the reflected signal.

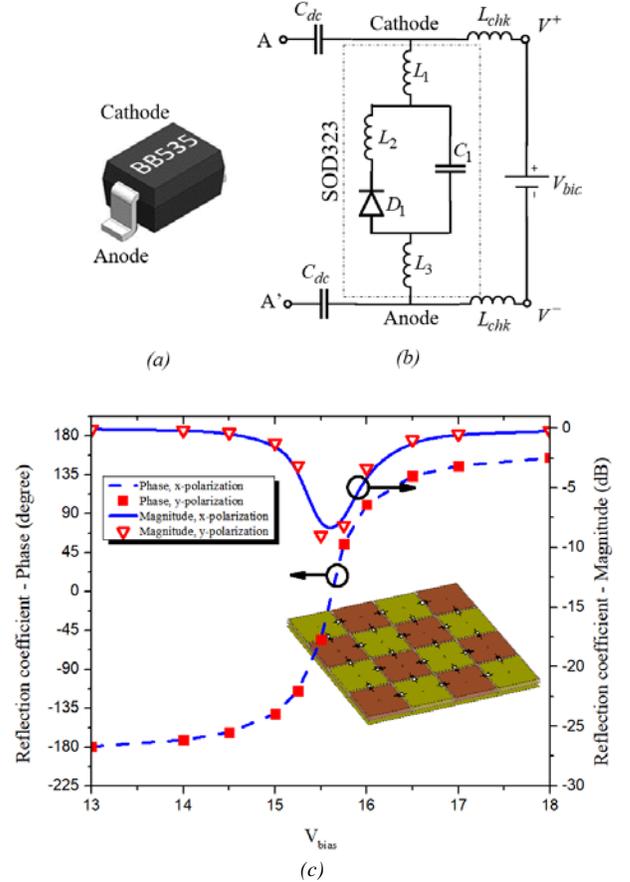

Fig. 4. (a) Varactor Infineon BB535 in package SOD323. (b) Loading circuit of the metasurface that takes into account all the parasitic elements of the varactor and its SPICE model D1. Filtering circuit elements $C_{dc}, L_{chk}$ act as a dc-block and low-pass filter, respectively. (c) Amplitude and phase of the reflection coefficient as a function of the polarization voltage $V_{bias}$ for both polarizations of the impinging electromagnetic field. The illumination frequency is $f_i = 1.5$ GHz.

### IV. TIME-VARYING METASURFACE

To verify the modulation-induced frequency shift of the reflected electromagnetic field with respect to the incident one, the temporal profile of the phase of the reflection coefficient exhibited by the metasurface should match the one analytically derived in Section II. Therefore, in this Section, we first retrieve the required temporal profile of the modulating signal in order to compensate the nonlinear phase of the reflection coefficient of the loaded metasurface with respect to the biasing voltage (Fig. 4c). Then, a full-wave

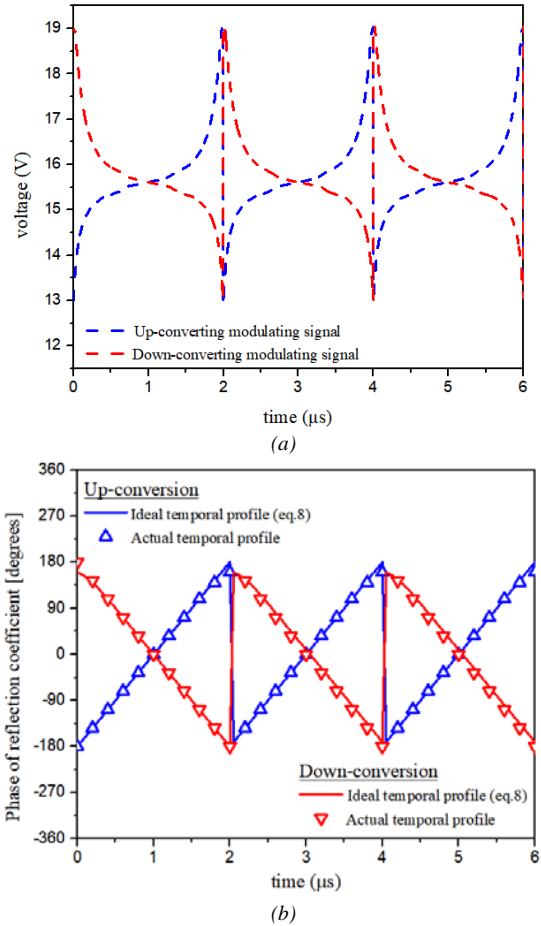

Fig. 5. (a) Driving voltage modulating signals at frequency $f_m = 500$ kHz for achieving both up- and down conversion (blue and red dashed line, respectively). (b) Comparison between the ideal time-dependent phase of the reflection coefficient from eq. (8) and the one exhibited by metasurface when the signals in Fig. 5a are applied.

simulation of the loaded time-modulated metasurface is performed.

*A. Design of the modulating voltage signal*

In Section II, eq. (8) describes analytically the required temporal profile of the complex reflection coefficient that enables a frequency shift from the incident frequency $f_i = \omega_i/2\pi$ to the reflected frequency $f_r = \omega_r/2\pi$. The reflection coefficient phase must exhibit a linearly varying temporal profile as a function of time, following the profile $\varphi(t) = (\omega_r - \omega_i)t$.

If the relationship between the phase of the reflection coefficient of the loaded metaurface and the biasing voltage were linear, a modulating voltage signal that linearly increase with time would be used to directly drive the varactors. However, from Fig. 4, it is clear that in the case of the metasurface loaded with varactors, such a relationship is strongly nonlinear, due mainly to the nonlinear response of the loaded metasurface with respect to the applied effective capacitance of the varactors and the intrinsic nonlinear response of the varactor itself. Therefore, the temporal profile of the driving modulating signal has to be selected in order to make the phase of the reflection coefficient of the metasurface linearly varying with time.

Let us assume to drive the varactors with a very low-frequency modulation signal with respect to the operative frequency of the metasurface, i.e., $f_m = 500$ kHz. The temporal profile of the modulating signal can be derived by mapping the proper value of the biasing voltage for each instant of time that permits the desired phase of the reflection coefficient at the same time. For the metasurface proposed in Section III, we report in Fig. 5 the retrieved modulating signals $V_{bias}(t)$ to be used for up-coversion (blue-dashed line - Fig. 5a) and down-conversion (red-dashed line – Fig. 5a). These voltage signals linearize the actual phase of the reflection coefficient exhibited by the time-varying metasurface as shown in Fig. 5b, where the comparison between the ideal time-dependent phase of the reflection coefficient from eq. (8) and the one exhibited by the metasurface when the signals in Fig. 5a are applied is reported.

In the following sub-section, the voltage signals of Fig. 5a are used for driving the loading varactors while the metasurface is illuminated by a monocromatic plane wave.

*B. Time-domain simulation of the time-varying metasurface*

The capability of the time-varying metasurface to up- and down convert the frequency of the incident field has been numerically verified though a co-simulation of CST Studio Suite [57] and Advanced Design System (ADS) [58]. The electromagnetic response has been evaluated by performing a time-domain transient simulation where the normally impinging illuminating field $\underline{E}_i$ at frequency $f_i = 1.5$ GHz and the driving voltage $V_{bias}(t)$ at freqeuncy $f_m = 500$ kHz are simultanously applied. The loading circuit used in the ADS simulation is topologically identical to the one used for characterizing the metasurface in Section III (Fig. 4b), but the biasing voltage $V_{bias}$ is now function of time according to the time-dependent proile shown in Fig. 5a and the filtering elements $C_{dc}$, $L_{chk}$ have been designed to ensure an efficient filtering of the low and high frequency components simulatenously present at the varactor terminals.

The time-domain reflected field has been evaluated and the corresponding Fourier transforms are reported in Fig. 7. When the driving voltage signals for up-conversion (blue dashed curve in Fig. 5a) is applied, the main frequency component of the reflected field is at the blue-shifted frequency $f_{r+} = f_i + f_m = 1.5005$ GHz, as shown in Fig. 6a. On the contrary, when the driving voltage signals for down-convertion (red dashed curve in Fig. 5a) is applied, the reflected field exhibits a main frequency component at $f_{r-} = f_i - f_m = 1.4995$ GHz, as shown in Fig. 6b.

Although the driving signal is designed to guarantee that the reflected spectra exhibit only the single frequency components at $f_{r\pm} = f_i \pm f_m$, the reflected energy spectra in Fig. 7 present some non-negligible frequency components at the incident frequency $f_i$ and $f_n = f_{r\pm} \pm nf_m$ with $n \neq 1$. This is maily caused by the divergence of the reflection coeffcient from the ideal response. Indeed, the amplitude of the reflection coefficient of the proposed metasurface is not always unitary in time as assumed in Section II, but it exhibits a periodical drop in corrisponcence of the zero phase reflection as reported in Fig. 4c. Therefore, its Fourier transform is not more a single pulse centered at $f_m = 500$ kHz as desired, but a set of harmonics, that are all convolved with the monochromatic incident field. Nevertheless, the proposed time-varying metasurface efficiently realizes an up- and down-frequency conversion of the illuminating signal, representing, to the authors' best knowledge, the first example of artificial Doppler effect, realized through an electrically thin time-modulated metasurface.

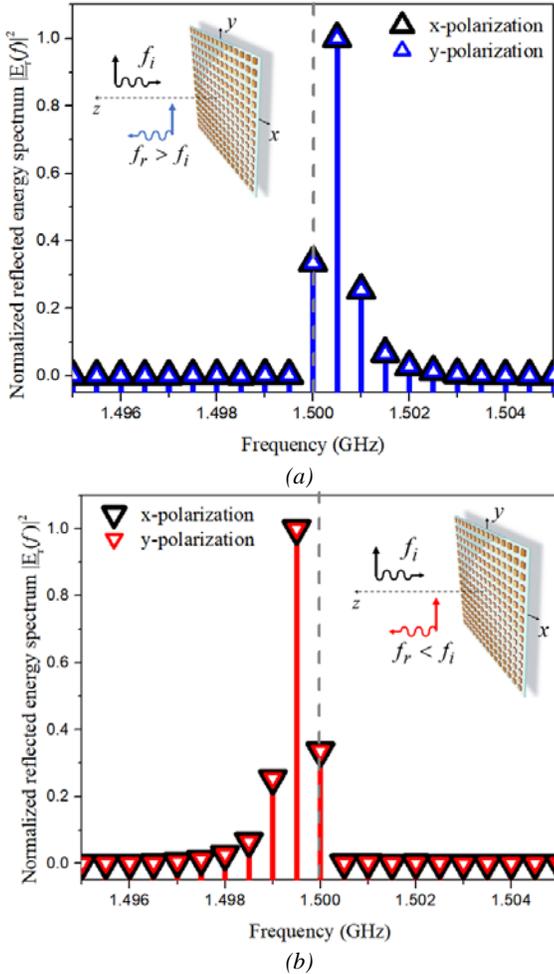

Fig. 6. Normalized reflected energy spectrum of (a) blue-shifted and (b) red-shifted reflected electromagnetic fields for both x- and y- polarizations.

V. CONCLUSION

In this paper, we have presented a design approach for time-modulated reflective metasurfaces, able to red- and blue-shift the frequency of the incident electromagnetic wave. We have analytically derived the necessary conditions for achieving the desired frequency shift. Based on the analysis, we have proposed a realistic time-varying metasurface, consisting of a properly designed and dynamically tuned high-impedance surface. Taking into account all the dissipative effects of the materials and the parasitic circuit elements, we have demonstrated that the proposed time-varying metasurface is able to frequency modulated the incident field "on-the-fly".

Such a capability may have a relevant impact in many applicative scenarios. For example, in the next-generation mobile telecommunication systems, where the proposed metasurface can be used to guarantee a continuous frequency hopping within ever smaller cells to maximize the channel capacity, or in Doppler sensors and radars, where the motion of a given object can be hidden by restoring in reflection the frequency of the incident electromagnetic field.